\documentclass{ws-ijmpa}
\usepackage[super,compress]{cite}
\usepackage{graphicx}
\usepackage{hyperref}
\usepackage{amsmath}
\usepackage{xcolor}
\usepackage{float}

\begin{document}

\title{Study of Quarkonium properties using SUSYQM method with baryonic chemical potential}

\author{Siddhartha Solanki, Manohar Lal, Rishabh Sharma, and Vineet Kumar Agotiya}
\address{Department of Physics, Central University of Jharkhand, Ranchi, India, 835 222.\\
Corresponding author id- {\bf agotiya81@gmail.com}}
\maketitle

\maketitle

\begin{abstract}
In this article, we employed the Quasi-particle debye mass at finite baryonic chemical potential which can be used in the medium modified heavy quark potential to solve the N-dimensional Schroedinger equation. The bound state solution of the Schroedinger equation using Cornell potential is obtained by Super-Symmetry Quantum Mechanics (SUSYQM) method. The thermodynamical properties of quark matter is calculated by using baryonic chemical potential ($\mu$). We found that the binding energy of quarkonia dissociates more with quasi-particle debye mass in comparison to non-perturbative and leading order debye mass. The medium modified form of potential (real part) has been used to study the thermodynamical properties of quark matter with different equation of states (EoS) (i.e., pressure, energy density and speed of sound) with $\mu$. The mass spectra of quarkonia has been also calculated in the N-dimensional space, and compared with the experimental data at $N$=$3$. We have also calculated the dissociation temperature ($T_{D}$) for the ground states of quarkonium using the dissociation criteria of thermal width.\\

\end{abstract}


\section{Introduction}
We know that the strong interaction occurring at the subatomic level are described by a well known theory of quantum Chromo-dynamics (QCD). This theory predicted that there is phase transition between the hadrons and the quark-gluon plasma by means of either at high temperature or at large baryon densities. This type of phase transition, which is supposed to occurred at the evolution of the early universe, can be recreated in the contemporary heavy ion collision like in the Brookhaven National Laboratory, USA and Large Hadron Collider experiment in the Switzerland. One of the remarkable property of the strongly interacting matter is that it behaves like a perfect fluid. So we can easily described this relativistic fluid by the relativistic hydrodynamics\cite{1,2,3,4,5}.\\ 
An important aspect of the hydrodynamics fluid is to describe the equation of states (EoS) which is related with the pressure, temperature, and local thermodynamic quantities. To describe the hydrodynamics, EoS plays crucial role. In the QCD phase transition, temperature and the baryon densities are the parameters which describes well the nuclear matter properties at subatomic levels. Lattice QCD theory is the first principle to explain the EoS of the nuclear matter and hence the QCD phase diagram in the T-$\mu$ plane, well described these EoS and hence the phase diagram. It has been found in\cite{6} that at $\mu=0$, there is cross-over and pseudo critical temperature, $T_{c}$ was obtained.\\
Lattic QCD at vanishing baryon density should in principle give an unambiguous result. Though QCD at finite chemical potential can be formulated on the lattice\cite{7}, whereas Monte Carlo techniques failed at non-zero baryon density. Schroedinger equation has played a crucial role in understanding the spectra of the quarkonia. This also describes the energy eigen-value, mass spectra of the quarkonium states and hence the thermodynamics properties by virtue of the partition function. From last three decades various mathematical models have been employed to solve the Schroedinger equation and used to describe the properties of the hot quark-gluon plasma. The  solution thus obtained may be approximate or exact. Keeping this view in mind various form of the potentials has been used to study the properties of the bound states (quarkonia) and there corresponds numerous method for solving Schroedinger equation such as Nikiforov Uvarov method\cite{8,9,10}. The AIM method provide complete solution for the harmonic oscillator potential and sextic enharmonic potential\cite{11}. Power series method\cite{12,13}, path integral method\cite{14}, Laplace transform method\cite{15,16} and analytic exact iteration method (AEIM)\cite{17} are some other useful method. The N-dimensional radial Schroedinger equation has been solved for various potential such as\cite{16} Mie Type potential\cite{18}, Cornell potential\cite{19,20} and extended Cornell potential\cite{17}. To study quarkonium properties\cite{21,22} modifying both part of potential (Coulombic and String) using hard thermal loop (HTL) dielectric permittivity for both isotropic and anisotropic media. So far, various attempts have been made to study the properties of the quarkonia at zero chemical potential of the thermal medium which mainly used leading order (LO) debye mass and numerical method to solve Schroedinger equation. Quarkonia properties has been also studied at finite temperature and finite chemical potential using LO debye mass in their calculations\cite{21}. N-dimension has been solved by\cite{23} for understanding the properties of the quarkonia using LO debye mass. Most of the recent studies\cite{24,25} mainly focus on properties of the QGP in the presence of finite temperature.\\
We used the SUSYQM method to obtained the solution of the N-dimensional Schroedinger equation. The SUSYQM method allow us to determine the eigen values and eigen function analytically for solvable potential\cite{26,27,28,29,30,31,32} by means of algebraic operator formulation. Here we take the medium modified form of the Cornell potential for studying the behavior of the quarkonia and hence its thermodynamical properties. Our present work is different from\cite{16,17,19,20}. In ways were we solve the Schroedinger equation by SUSYQM method. The point is that we use Quasi-particle (QP) debye mass with baryonic chemical potential in our present work which has been not reported in the studies to best of our knowledge. Another aspect of our work is that we study the thermodynamical properties using Quasi-particle debye mass with baryonic chemical potential, which makes it novel. However, some studies\cite{33}, consider thermodynamic properties with Rose morse potential to solve Schroedinger equation by using AEIM method. Here, medium modified form of the potential has been used to study the properties of quarkonia. The QP debye mass which is used in this work depends on the temperature and baryonic chemical potential.\\  
The present work is organized in the following manner: The energy eigen value has been obtained by SUSYQM method in the section II by using modified form of Heavy quark potential. In section III, a brief description about the imaginary part of the potential is given. Section IV, describes about the quasi particle debye mass in the Hot and Dense medium. Whereas in section V, calculation of various thermodynamical properties of matter with baryonic chemical potential using the formulation of equation of states (EoS). Section VI, briefly describes about the mass spectra of quarkonium state in $N$-dimensional space. Finally, results and conclusion of the work are given in section VII.

\section{Solution of Schroedinger equation and energy eigen values.}
To study the properties of charmonium and bottomonium S-wave in the presence of finite baryonic chemical potential in N-dimensional space, we start with the solution of N-dimensional Schroedinger equation by using SUSYQM method. The concept of SUSYQM mechanics has been widely used in the elementary particle physics since $1970$, and Witten\cite{34} was the first to study the SUSY model in the quantum field theory.\\ 
The radial Schroedinger equation in the N-dimensional space for the two interacting quarkonia with asymmetric potential\cite{35,36} can take the forms as:
\begin{eqnarray}
\label{eq1}
\left \{ \frac{\partial^2 }{\partial r^2} + \frac{N-1}{r}\frac{\partial }{\partial r}+2 \mu_{Q \bar{Q}} \left ( E_{nl}-V(r)-\frac{l(l+N-2)}{2 \mu_{Q\bar{Q}}r^{2}} \right )\right \}\psi(r)=0
\end{eqnarray}
Where, $\mu_{Q \bar{Q}} $, l and N  denotes reduced mass of the system, the orbital momentum quantum number, and the Dimensionality number respectively. $E_{nl}$ represent the energy eigen value. We define $\psi (r)$, the wave function as 
$\psi (r)= r^{\frac{1-N}{2}} R(r)$.
Using this wave function the Eq.(\ref{eq1}) becomes, 
\begin{figure*}
\includegraphics[height=5cm,width=6cm]{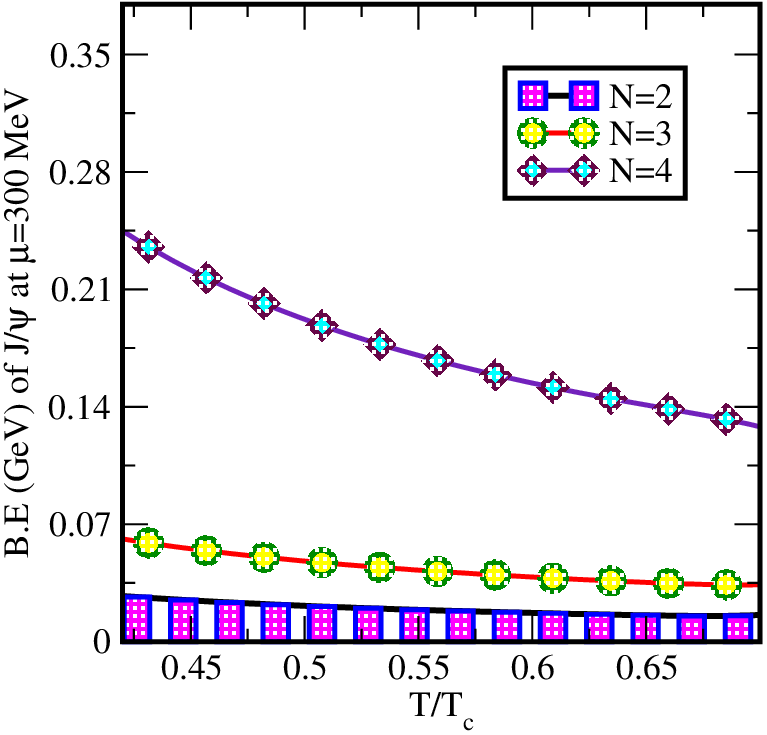}
\hspace{3mm}
\includegraphics[height=5cm,width=6cm]{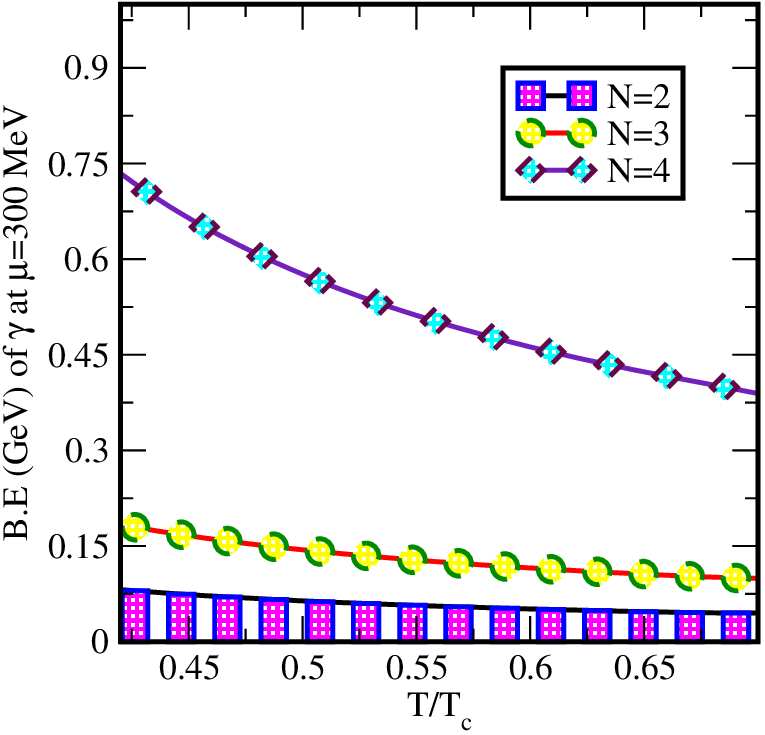}
\caption{Dependence of $J/\psi$ binding energy with temperature (left panel) and dependence of $\Upsilon$ binding energy with
temperature (right panel) at different values of $N$ but the value of $\mu$ is fixed.}
\label{fig.1}
\vspace{1mm}
\end{figure*}
\begin{eqnarray}
\label{eq2}
\left \{ \frac{\partial^2 }{\partial r^2}+2 \mu_{Q \bar{Q}} \left ( E_{nl}-V(r)-\frac{[(l+\frac{N-2}{2})^2-\frac{1}{4}]}{2 \mu_{Q\bar{Q}}r^{2}} \right )\right \}R(r)=0
\end{eqnarray}
The potential we use in the present work can be formed in\cite{22,Agotiyaprd} which is as follows:
\begin{eqnarray}
\label{eq3}
V(r)=\left ( \frac{2\sigma }{m_{D}^{2}}-\alpha \right )\frac{e^{-m_{D}r}}{r}-\frac{2\sigma }{m_{D}^2 r}+\frac{2\sigma }{m_{D}}-\alpha m_{D}
\end{eqnarray}
Where, $\alpha$ is the two loop coupling constant depending on temprature and $\sigma$ is the string constant which is taken $0.184$ $GeV^{2}$ and  $m_{D}$ is the temperature and baryonic chemical potential dependent QP Debye mass. By putting Eq.(\ref{eq3}) into Eq.(\ref{eq2}), we get,
\begin{multline}
\label{eq4}
\frac{\partial^2 }{\partial r^2}+2 \mu_{Q \bar{Q}}\\ \left ( E_{nl}-\left (\left ( \frac{2\sigma }{m_{D}^{2}}-\alpha \right )\frac{e^{-m_{D}r}}{r}-\frac{2\sigma }{m_{D}^2 r}+\frac{2\sigma }{m_{D}}-\alpha m_{D}\right )-\frac{[(l+\frac{N-2}{2})^2-\frac{1}{4}]}{2 \mu_{Q\bar{Q}}r^{2}} \right )R(r)=0
\end{multline}
The following approximation\cite{26} has been used for the simplicity of $\frac{1}{r}$ terms:
\begin{eqnarray}
\label{eq5}
\frac{1}{r}\approx \frac{m_{D}}{(1-e^{(-m_{D}r)})}
\end{eqnarray}
Now, By substitution of Eq.(\ref{eq5}) into Eq.(\ref{eq4}), a second-order Schroedinger equation is obtained:
\begin{eqnarray}
\label{eq6}
-\frac{\partial^2 R_{nl}(r) }{\partial r^2}+V_{eff}(r)R_{nl}(r)=\tilde{E}_{nl}R_{nl}(r)
\end{eqnarray}
Where,
\begin{eqnarray}
\label{eq7}
V_{eff}(r)=\frac{Ae^{-2m_{D}r}+Be^{-m_{D}r}+C}{(1-e^{-m_{D}r})^2}
\end{eqnarray}
with $A=0$, $B=2\mu_{Q \bar{Q}}\alpha m_{D}$ and $C=m_{D}^{2}\left [ \left ( l+\frac{N-2}{2} \right ) -\frac{1}{4}\right ]$-$2\mu_{Q \bar{Q}}\alpha m_{D}$. From Eq.(\ref{eq7}), The effective energy can be written as:
\begin{eqnarray}
\label{eq8}
\tilde{E}_{nl} =2\mu_{Q \bar{Q}} E_{nl}
\end{eqnarray}
The ground state wave function in SUSYQM can be defined as in\cite{26,37,38,39}:
\begin{eqnarray}
\label{eq9}
R_{0,l}(r)=exp\left ( -\int W(r)dr \right )
\end{eqnarray}
In Eq.(\ref{eq9}) the integrand W(r) is the superpotential and the Hamiltonian of the system which is combination of raising and lowering operators as defined in\cite{26,37,38,39}:
\begin{eqnarray}
\label{eq10}
H_{-}=\hat{A}^{+}\hat{A}=-\frac{\partial^2 }{\partial r^2}+V_{-}(r) 
\end{eqnarray}
and
\begin{eqnarray}
\label{eq11}
H_{+}=\hat{A}\hat{A}^{+}=-\frac{\partial^2 }{\partial r^2}+V_{+}(r) 
\end{eqnarray}
The lowering and raising operator for the W(r) defined as:
\begin{eqnarray}
\label{eq12}
\hat{A}^{+}=-\frac{\partial }{\partial r}-W(r)
\end{eqnarray}
and
\begin{eqnarray}
\label{eq13}
\hat{A}=\frac{\partial }{\partial r}-W(r)
\end{eqnarray}
Now, the value of partner Hamiltonian thus obtained are as follows:
\begin{eqnarray}
\label{eq14}
V_{\pm }(r)=W^{2}(r)\mp W^{'}(r)
\end{eqnarray}
In Eq.(\ref{eq14}) $W^{2}(r)\mp W^{'}(r)$ is the associated Riccati equation in SUSYQM and is of the form given below:
\begin{eqnarray}
\label{eq15}
W^{2}(r)\mp W^{'}(r)=V_{eff}(r)-\tilde{E}_{0,J} 
\end{eqnarray}
The super potential with the Eq.(\ref{eq7}) now takes the form:
\begin{eqnarray}
\label{eq16}
W(r)=\frac{Pe^{-m_{D}r}}{(1-e^{-m_{D}r})}+Q 
\end{eqnarray}
Putting Eq.(\ref{eq7}) and Eq.(\ref{eq14}) into Eq.(\ref{eq12}), we obtain the following sets of expressions:
\begin{eqnarray}
\label{eq17}
P=\frac{-m_{D}\pm \sqrt{m_{D}^{2}+4(A+B+C)}}{2}
\end{eqnarray}
\begin{eqnarray}
\label{eq18}
Q=\frac{P^{2}+C-A}{2P}
\end{eqnarray}
\begin{eqnarray}
\label{eq19}
\tilde{E}_{0,l}=C-Q^{2}
\end{eqnarray}
The two Hamiltonians partner can be written as in\cite{26,37,38,39}:
\begin{multline}
\label{eq20}
V_{-}(r)=W^{2}(r)-\frac{\mathrm{d} W}{\mathrm{d} r}=\frac{P(P-m_{D})e^{-2m_{D}r}}{(1-e^{-m_{D}r})^{2}}\\+\frac{2P\left ( \frac{P^{2}+C-A}{2P} \right )e^{-m_{D}r}}{(1-e^{-m_{D}r})}+\left ( \frac{P^{2}+C-A}{2P} \right )^{2}
\end{multline}
\begin{multline}
\label{eq21}
V_{+}(r)=W^{2}(r)+\frac{\mathrm{d} W}{\mathrm{d} r}=\frac{P(P+m_{D})e^{-2m_{D}r}}{(1-e^{-m_{D}r})^{2}}\\+\frac{2P\left ( \frac{P^{2}+C-A}{2P} \right )e^{-m_{D}r}}{(1-e^{-m_{D}r})}+\left ( \frac{P^{2}+C-A}{2P} \right )^{2}
\end{multline}
The following relationship is satisfied by the two potential partner Eq.(\ref{eq20}) and Eq.(\ref{eq21}):
\begin{eqnarray}
\label{eq22}
V_{+}(r,\rho _{o})=V_{-}(r,\rho _{1})+R(\rho _{1})
\end{eqnarray}
With $\rho_{o}$=$P$ and $\rho_{i}$ is a function of $\rho_{o}$, i.e. $\rho_{1}$=$f(\rho_{o})=\rho_{o}-m_D$. Therefore, $\rho_{n}$=$\rho_{o}$-$nm_{D}$.
Thus we can see that the shape invariance holds via a mapping of the form $P$ tends to $P$-$m_{D}$. from Eq.(\ref{eq20}), we get:
\begin{multline}
\label{eq23}
R(a_{1})=\left ( \frac{(\rho _{0})^2 +C-A}{2(\rho _{0})} \right )^{2}-\left ( \frac{(\rho _{1})^2 +C-A}{2(\rho _{1})} \right )^{2}\\
R(a_{2})=\left ( \frac{(\rho _{1})^2 +C-A}{2(\rho _{1})} \right )^{2}-\left ( \frac{(\rho _{2})^2 +C-A}{2(\rho _{2})} \right )^{2}\\
.\\
.\\
.\\
.\\
R(a_{n})=\left ( \frac{(\rho _{n-1})^2 +C-A}{2(\rho _{n-1})} \right )^{2}-\left ( \frac{(\rho _{n})^2 +C-A}{2(\rho _{n})} \right )^{2}
\end{multline}
The total energy eigen-values for the system is obtained as follows:
\begin{eqnarray}
\label{eq24}
\tilde{E}_{nl}=\tilde{E}_{nl}^{-}+\tilde{E}_{0,l}
\end{eqnarray}
Where,
\begin{eqnarray}
\label{eq25}
\tilde{E}_{nl}^{-}=\sum_{k=1}^{n}R(a_{k})=\left ( \frac{(\rho _{0})^2 +C-A}{2(\rho _{0})} \right )^{2}-\left ( \frac{(\rho _{n})^2 +C-A}{2(\rho _{n})} \right )^{2}  
\end{eqnarray}
Using Eq.(\ref{eq16}), Eq.(\ref{eq17}) and Eq.(\ref{eq23}) in Eq.(\ref{eq22}) we obtain: 
\begin{multline}
\label{eq26}
\tilde{E}_{nl}=-\left[\frac{ \left( \frac{-m_{D}+\sqrt{m_{D}^{2}+4 m_{D}^{2}\left [ \left ( l+\frac{N-2}{2} \right )^{2}-\frac{1}{4} \right ]}}{2}-nm_{D} \right )^{2}+\left \{m_{D}^{2}\left [ \left ( l+\frac{N-2}{2} \right )^{2}-\frac{1}{4} \right ]-m_{Q}\alpha m_{D}\right \}}{2\left ( \frac{-m_{D}+\sqrt{m_{D}^{2}+4m_{D}^{2}\left [ \left ( l+\frac{N-2}{2} \right )^{2}-\frac{1}{4} \right ]}}{2}-nm_{D} \right )} \right ]^{2}\\+\left \{m_{D}^{2}\left [ \left ( l+\frac{N-2}{2} \right )^{2}-\frac{1}{4} \right ]-m_{Q}\alpha m_{D}\right \}
\end{multline}
\begin{figure*}
\includegraphics[height=5cm,width=6cm]{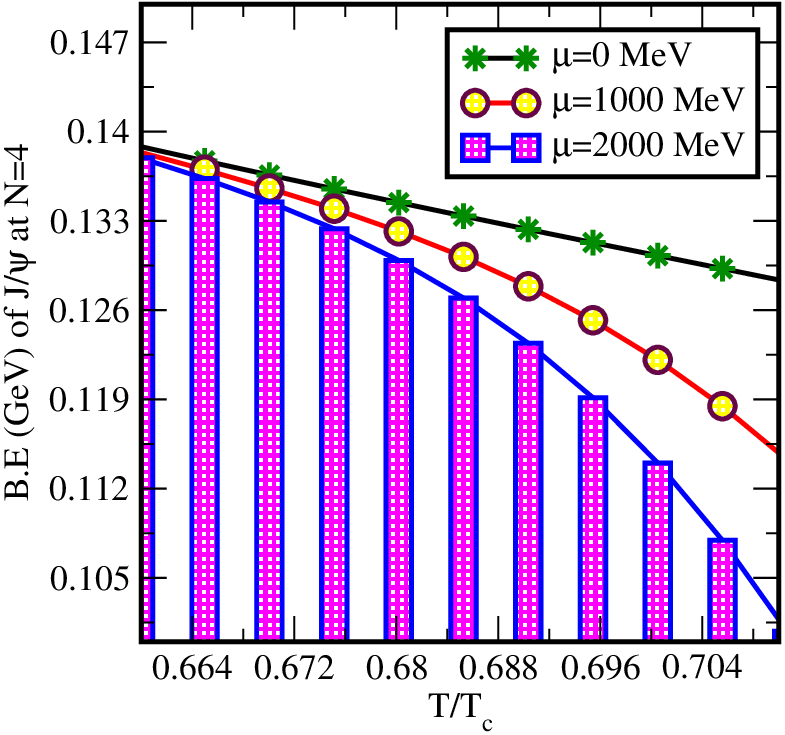}
\hspace{3mm}
\includegraphics[height=5cm,width=6cm]{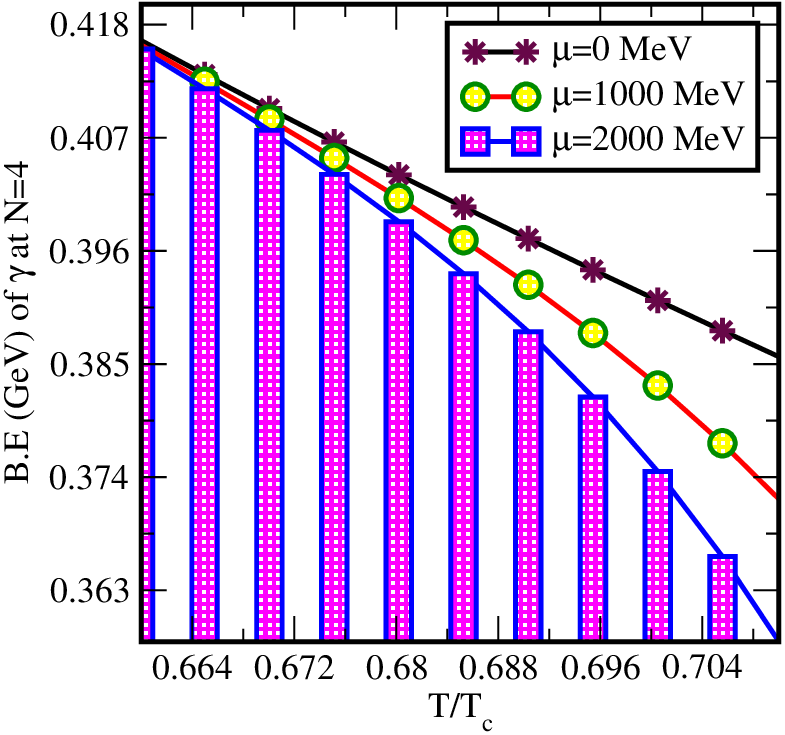}
\caption{Dependence of $J/\psi$ binding energy with $T/T_C$ (left panel) and dependence of $\Upsilon$ binding energy with
$T/T_c$ (right panel) at different values of $\mu$ but the value of $N$ is fixed.}
\label{fig.2}
\vspace{1mm}
\end{figure*}
Where, $m_{Q}$ is the mass of quarkonium. The energy spectrum for the quarkonium in the N-dimensional space obtained by substituting Eq.(\ref{eq8}) in Eq.(\ref{eq26}) is given below:
\begin{multline}
\label{eq27}
{E}_{nl}=-\frac{1}{m_{Q}}\left[\frac{ \left( \frac{-m_{D}+\sqrt{m_{D}^{2}+4 m_{D}^{2}\left [ \left ( l+\frac{N-2}{2} \right )^{2}-\frac{1}{4} \right ]}}{2}-nm_{D} \right )^{2}+\left \{m_{D}^{2}\left [ \left ( l+\frac{N-2}{2} \right )^{2}-\frac{1}{4} \right ]-m_{Q}\alpha m_{D}\right \}}{2\left ( \frac{-m_{D}+\sqrt{m_{D}^{2}+4m_{D}^{2}\left [ \left ( l+\frac{N-2}{2} \right )^{2}-\frac{1}{4} \right ]}}{2}-nm_{D} \right )} \right ]^{2}\\+\frac{1}{m_{Q}}\left \{m_{D}^{2}\left [ \left ( l+\frac{N-2}{2} \right )^{2}-\frac{1}{4} \right ]-m_{Q}\alpha m_{D}\right \}
\end{multline}

\section{Imaginary Part of the potential}
Heavy quark complex potential contains both the real and the imaginary part which were obtained by using the gluon self energy which in turn is responsible for both the Debye screening and the Landau damping respectively. However, Debye screening obtained by using both the retard and self-energy propagator. Whereas the static limit of the symmetric self-energy has been used for calculating the imaginary part of the potential. The imaginary part of the potential has its importance while studying the threshold enhancement of the bound state or the thermal width of the charmonium and bottomonium resonances. This thermal width in spectral function further used to determine the dissociation temperature of the quarkonium resonances. From the studies\cite{40,41}, the dissociation of the quarkonium states takes place when thermal width becomes equal to the twice  of the binding energy.
The symmetric propagator for the imaginary part is given below:
\begin{equation}
\label{eq28}
ImD_F^{00}(0,p)=\frac{-2\pi{T} {m_D^2}}{p\left(p^2+m_D^2\right)^2}
\end{equation}
Eq.\ref{eq28} gives the imaginary part of the dielectric function as:
\begin{equation}
\label{eq29}
\epsilon^{-1}(p)=-\pi{T}{m_D^2}\ \frac{p^2}{p\left(p^2+m_D^2\right)^2}
\end{equation}
Thus, we obtained the imaginary part of the potential using the following equation:
\begin{equation}
\label{eq30}
V(r,T,\mu)=\int{\frac{d^3k}{\left(2\pi\right)^\frac{3}{2}}\left(e^{i.pr}-1\right)\frac{V(p)}{\epsilon(p)}}
\end{equation}
This implies
\begin{multline}
\label{eq31}
ImV\left(r,T,\mu\right)=\int{\frac{d^3k}{\left(2\pi\right)^\frac{3}{2}}\left(e^{i.pr}-1\right)\left(-\sqrt{\frac{2}{\pi}}\frac{\alpha}{p^2}-\frac{4\sigma}{\sqrt{\pi}{p^4}}\right)}\\\times p^{2}\left( {\frac{-\pi^{2}{T} {m_D^2}}{p\left(p^2+m_D^2\right)^2}} \right)
\end{multline}
After performing the integration of the above Eq.(\ref{eq31}), the contribution due to coulomb and the string term becomes:
\begin{multline}
\label{eq32}
ImV\left(r,T,\mu\right)= -2{\alpha}{T}\int_{0}^{\infty}\\{\frac{dz}{{(z^2+1)}^2}\left(1-\frac{\sin{z}}{z\hat{r}}\right)+\frac{4\sigma{T}}{m_D^2}\int_{0}^{\infty}{\frac{dz}{{(z^2+1)}^2}\left(1-\frac{\sin{z\hat{r}}}{z\hat{r}}\right)}}
\end{multline}
Where z=$\frac{P}{m_D}$. This can be further simplified  as:
\begin{equation}
\label{eq33}
ImV\left(r,T,\mu\right)=\-{\alpha}{T}\phi_0\left(\hat{r}\right)+\frac{2\sigma{T}}{m_D^2}\psi_0\left(\hat{r}\right)
\end{equation}
Where  $\phi_0\left(\hat{r}\right)=-{\alpha}{T}\left(\frac{{\hat{r}}^2}{9}\left(-4+3\gamma_E+ 3log(\hat{r})\right)\right)$\
~\\
 and 
 ~\\
 $\psi_0\left(\hat{r}\right)=-\frac{{\hat{r}}^2}{6}\ +\left(\frac{-107+60\gamma_E+60log(\hat{r})}{3600}\right){\hat{r}}^4+O({\hat{r}}^5)$
~\\
For the limit $\hat{r}<<1$, we have
\begin{equation}
\label{eq34}
ImV\left(r,T,\mu\right)=-T\left(\frac{\alpha{\hat{r}}^2}{3}+\frac{\sigma{\hat{r}}^4}{30m_D^2}\right)log\left(\frac{1}{\hat{r}}\right)
\end{equation}
Now, by using the Eq.(\ref{eq34}) we calculate the thermal width $\Gamma$ of the resonance state (1S). This can be done by folding with unperturbed (1S) Coulomb wave function and is given by:
\begin{equation}
\label{eq35}
\Gamma=\left(\frac{4T}{\alpha{m_Q^2}}+\frac{12\sigma{T}}{\alpha^2m_Q^4}\right)m_D^2log\frac{\alpha{m_Q}}{2m_D}
\end{equation}
\begin{figure*}
\centering
\includegraphics[height=6cm,width=5.5cm]{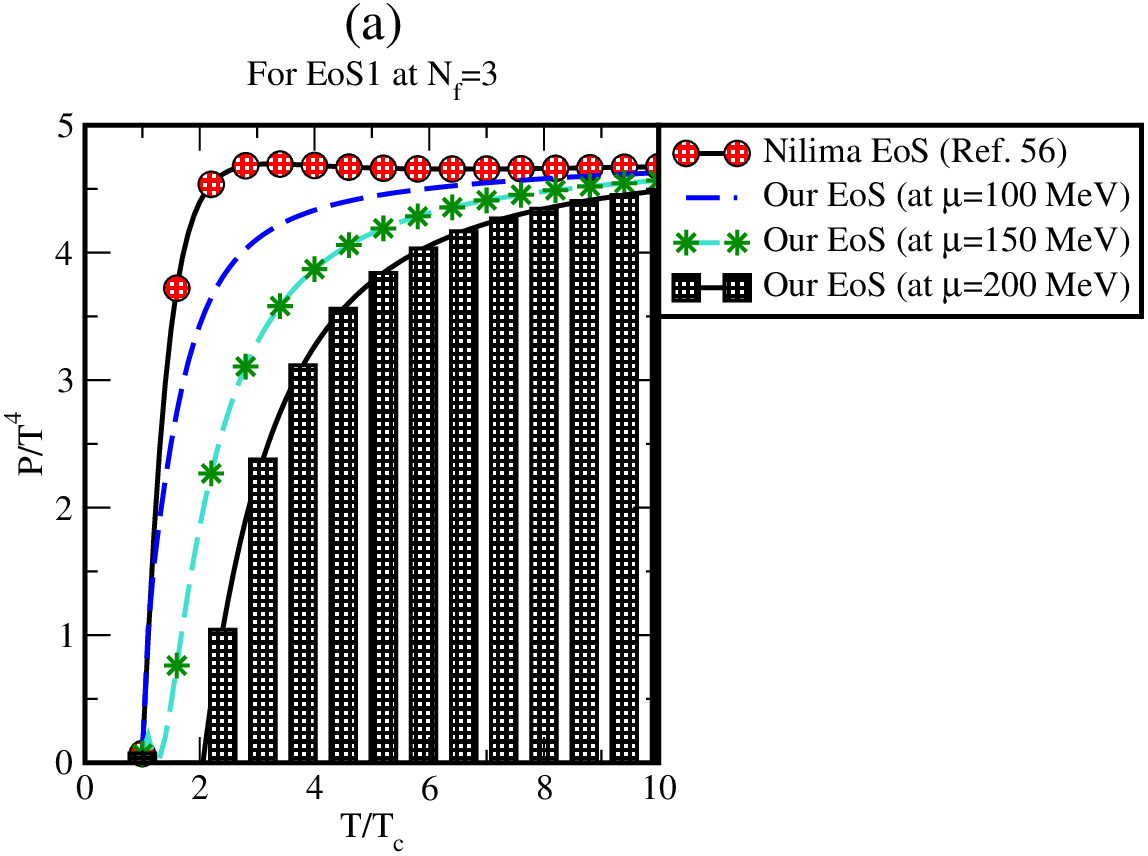}
\hspace{3mm}
\includegraphics[height=6cm,width=5.5cm]{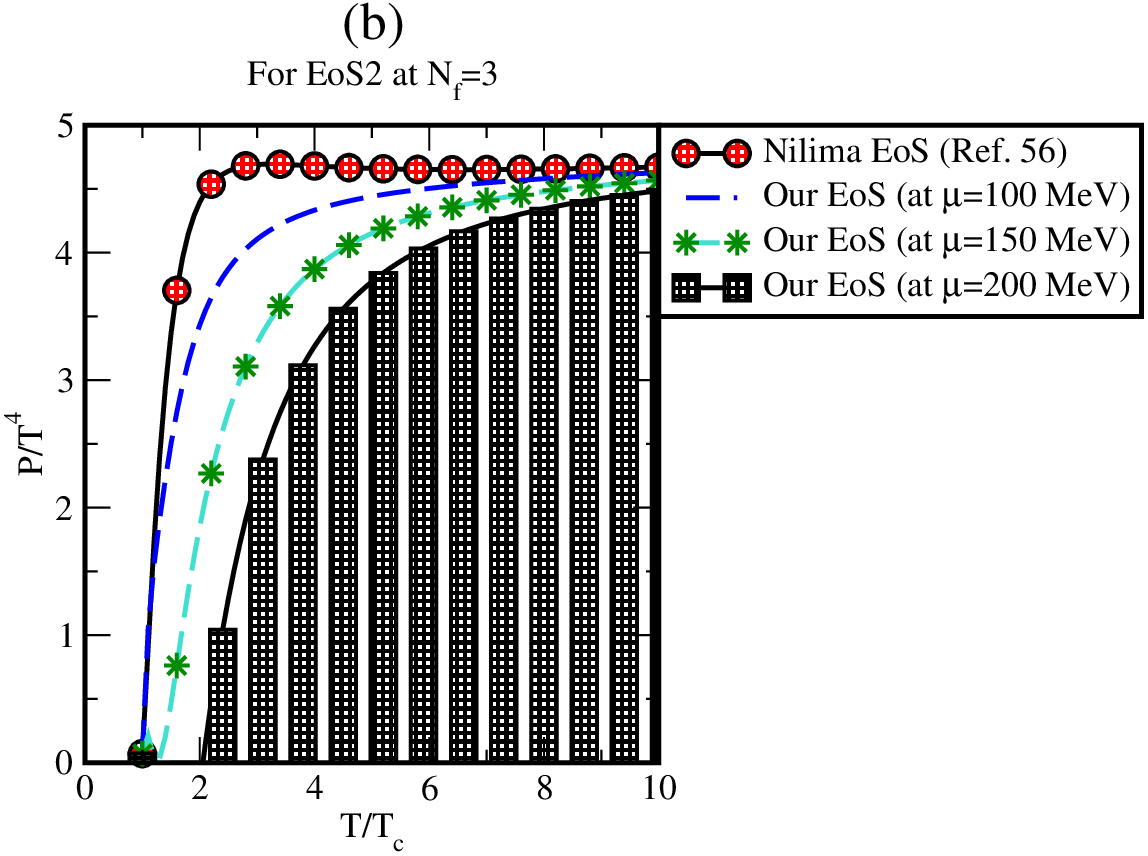}
\vspace{2cm}
\caption{Plot of $\frac{P}{T^{4}}$ as a function of $\frac{T}{T_{c}}$ for 3-flavor QGP for EoS1 (fig.3(a)) and for EoS2 (fig.3(b)).}
\label{fig.3}
\vspace{3cm}
\end{figure*}
\begin{figure*}
\centering
\includegraphics[height=6cm,width=5.5cm]{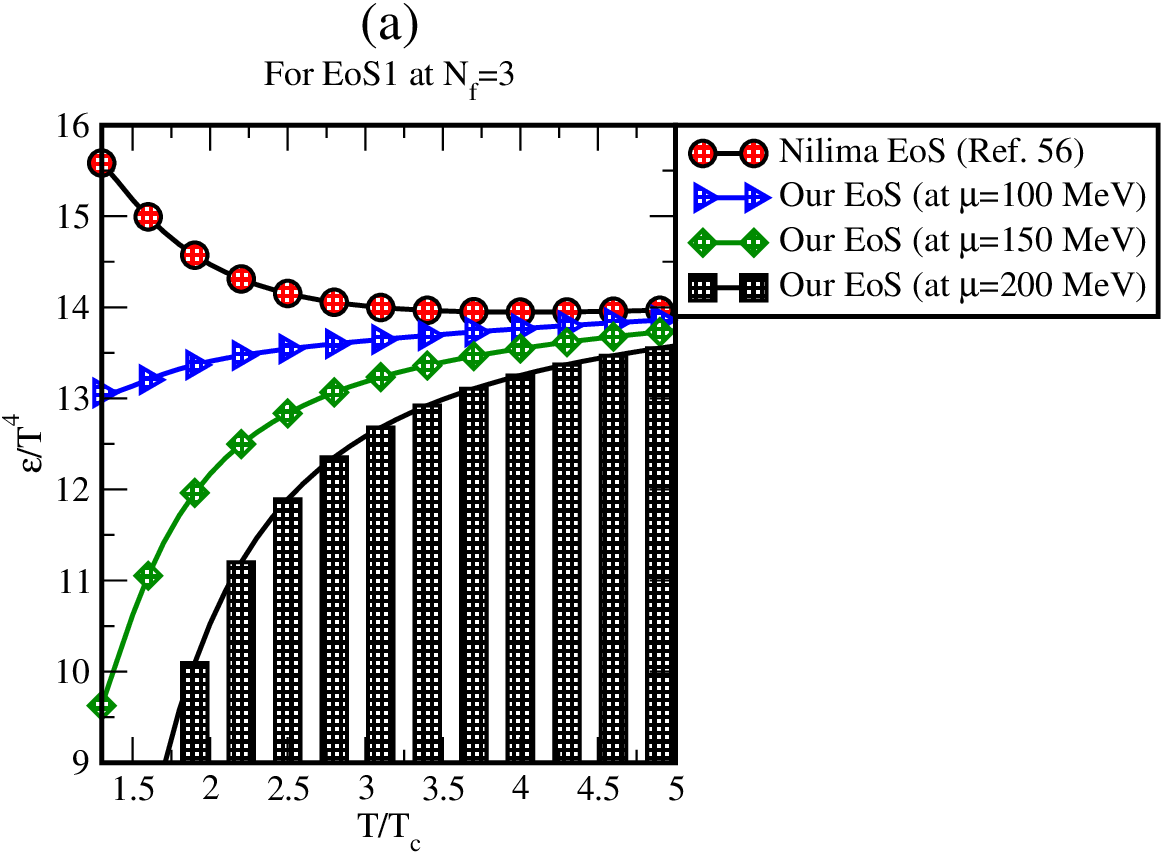}
\hspace{3mm}
\includegraphics[height=6cm,width=5.5cm]{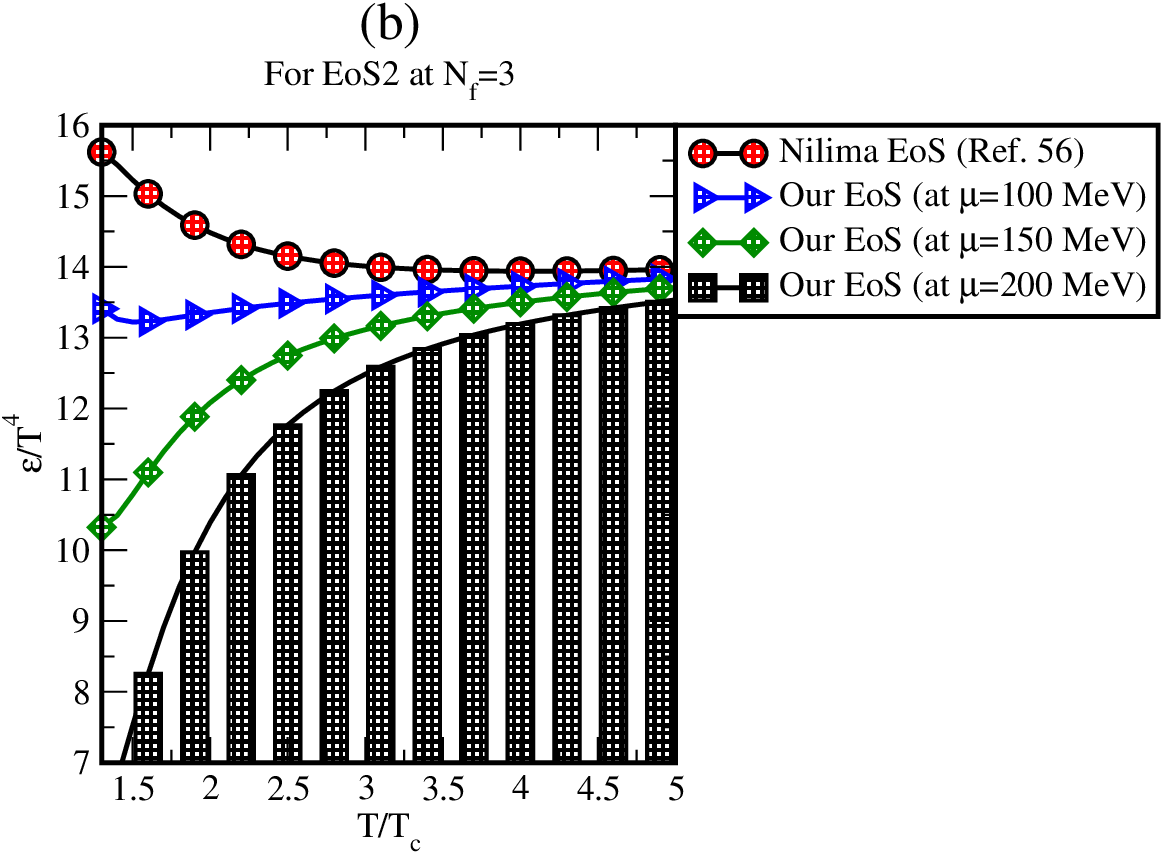}
\vspace{2cm}
\caption{Plot of $\frac{\epsilon}{T^{4}}$ as a function of $\frac{T}{T_{c}}$ for 3-flavor QGP for EoS1 (fig.4(a)) and for EoS2 (fig.4(b)).}
\label{fig.4}
\vspace{2cm}
\end{figure*}

\section{Quasi-particle picture of Debye mass in hot QCD}

Debye mass is an important quantity to study the quark gluon plasma, thus it is necessary to define the Debye mass in the perturbative regime (high temperature) and is defined in\cite{ARebhan} as the pole of the static propagator component of the gluon self energy. The Debye mass thus obtained is gauge independent, this is due to the fact that the gluon self energy have not any gauge choice. The studies of Nieta\cite{EBraaten}, Arnold and Yaffe\cite{PArnold} pointed out the Debye mass as a pole of the gluon propagator that does not hold good. In lattice QCD, the Debye mass itself creates a difficulty this is because of the fact that in QCD the electric field correrelator are not gauge invariant. This problem was sort out by the various studies and one can find the  details in\cite{KKajantie,SNadkarni}, Burnier and 
Rothkopf\cite{YBurnier}.\\
To describe the QCD EoS in terms of non-interacting quasi-partons (quasi-gluons and quasi-quarks) several attempts has been made. The quasiparton could be defined as the excitations of the interacting quarks and gluons. Number of models have been used to describe the quasi-partons such as, effective mass model\cite{VGoloviznin,peshier}, effective mass model with polykov loop\cite{MDElia}, model based on PNJL and NJL\cite{ADumitru} and effective fugacity model\cite{VChandra7,VChandra9}. In quantum chromodynamics, to describe the non-linear behavior of the quark gluon plasma near the transition point, we use quasi particle model, a phenomenological model. Here we assumed system of interacting massless quarks and gluon as an ideal gas of massive non interacting quasi particle. The quasi-particles has temperature dependent mass which arises due to the interacting quarks and gluons with surrounding medium. The quasi particle retain the quantum number of the quarks and gluons\cite{PKSrivastava}.
In our calculation, we use the Debye mass $m_D$ for the full QCD case and is given by:
\begin{multline}
\label{eq36}
m^2_D\left(T\right) = g^2(T) T^2\\ \bigg[\bigg(\frac{N_c}{3}\times\frac{6 PolyLog[2,z_g]}{\pi^2}\bigg)+\bigg(\frac{\hat{N_f}}{6}\times\frac{-12 PolyLog[2,-z_q]}{\pi^2}\bigg)\bigg]
\end{multline}
and 
\begin{eqnarray}
\label{eq37}
\hat{N_f} = \bigg(N_f +\frac{3}{\pi^2}\sum\frac{\mu_{i}^2}{T^2}\bigg)
\end{eqnarray}
Here, $\mu_{i}$ is quark chemical potential and $g (T)$ is the temperature dependent two loop running coupling constant, $N_c$=$3$ ($SU(3)$) and $N_f$ is the number of flavor, the function $PolyLog[2,z]$ having form, $PolyLog[2,z]$=$\sum_{k=1}^{\infty} \frac{z^k}{k^2}$ and $z_g$ is the quasi-gluon effective fugacity and $z_q$ is quasi-quark effective fugacity. These distribution functions are isotropic in nature,
\begin{eqnarray}
\label{eq38}
f_{g,q}=\frac{z_{g,q}exp(-\beta p)}{\left (1\pm z_{g,q}exp(-\beta p)  \right )}
\end{eqnarray}
Where, $g$ stands for  quasi-gluons and $q$ for quasi-quarks.
These fugacities should not be confused with any conservation's law (number conservation) and have merely been introduced to encode all the interaction effects at high temperature QCD. Both $z_g$ and $z_q$ have a very complicated temperature dependence and asymptotically reach to the ideal value unity~\cite{VChandra9}.
The temperature dependence $z_g$ and $z_q$ fits well to the form given below,
\begin {equation}
\label{eq39}
z_{g,q}=a_{q,g}\exp\bigg(-\frac{b_{g,q}}{x^2}-\frac{c_{g,q}}{x^4}-\frac{d_{g,q}}{x^6}\bigg).
\end {equation}
(Here $x=T/T_c$ and $a$, $b$, $c$ and $d$ are fitting parameters), for both EoS$1$ and EoS$2$. Here, EoS$1$ is the $O (g^5)$ hot QCD~\cite{PArnold} and EoS$2$ is the $O (g^6\ln(1/g)$ hot QCD EoS~\cite{KKajantie} in the quasi-particle description~\cite{VChandra7,VChandra9} respectively.
Now, the expressions for the Debye mass can be rewritten in terms of effective charges for the quasi-gluons and quarks for pure gauge as:
\begin{equation}
\label{eq40}
m^2_D\left(T\right)= Q^2_g T^2\frac{N_c}{3}
\end{equation}
and for the full QCD case in terms of baryonic chemical potential is,
\begin{eqnarray}
\label{eq41}
m^2_D\left(T,\mu\right)=T^2\left \{ \bigg( \frac{N_c}{3} Q^2_g\bigg)+\left \{ \bigg[\frac{N_f}{6}+\frac{1}{2\pi^2}\bigg(\frac{\mu^2}{9T^2}\bigg)\bigg]Q^2_q\right \}\right \}
\end{eqnarray}
Where, $\mu$ is the baryonic chemical potential and is equal to three times of quark chemical potential $\mu_{i}$. $Q_g$ and $Q_q$ are effective charges given by the equations:
\begin{eqnarray}
\label{eq42}
Q^2_g&=&g^2 (T) \frac{6 PolyLog[2,z_g]}{\pi^2}\nonumber\\
Q^2_q&=&g^2 (T)  \frac{-12 PolyLog[2,-z_q]}{\pi^2}.
\end{eqnarray}
\begin{figure*}
\centering
\includegraphics[height=6cm,width=5.5cm]{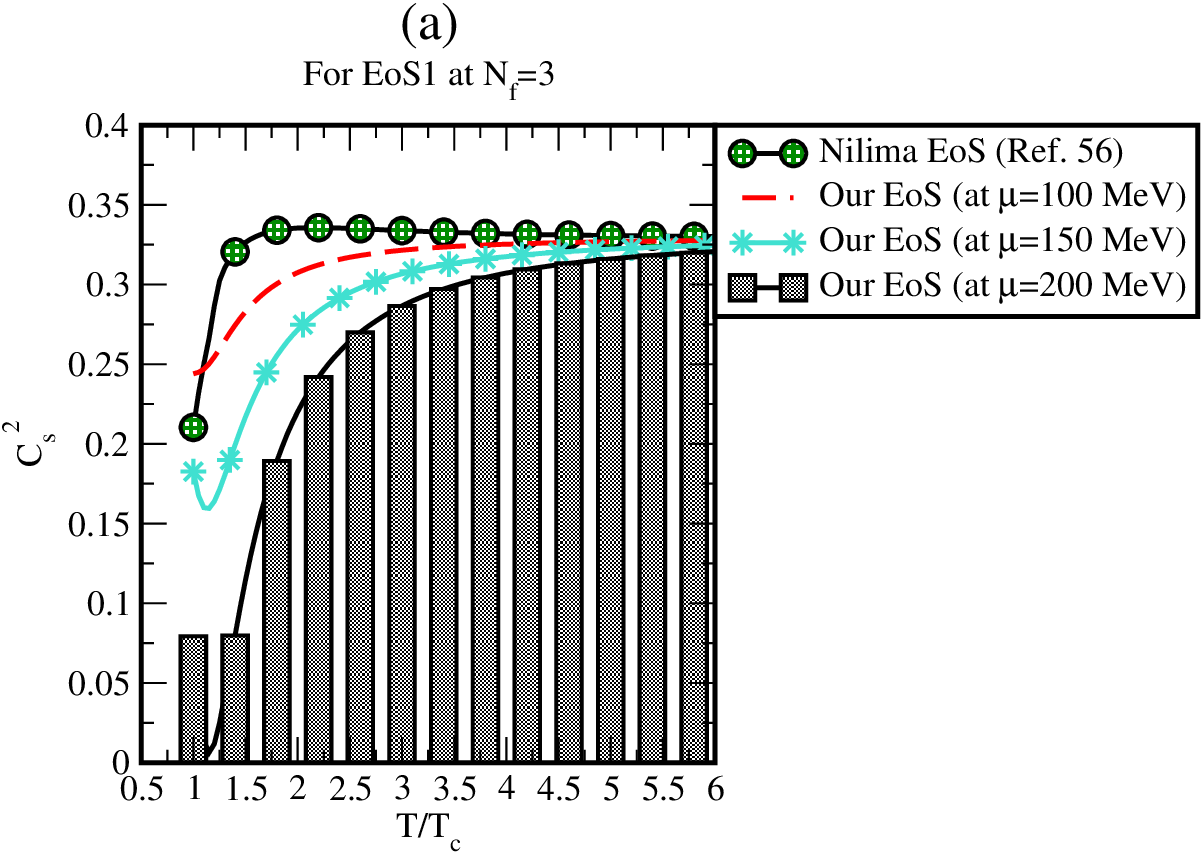}
\hspace{3mm}
\includegraphics[height=6cm,width=5.5cm]{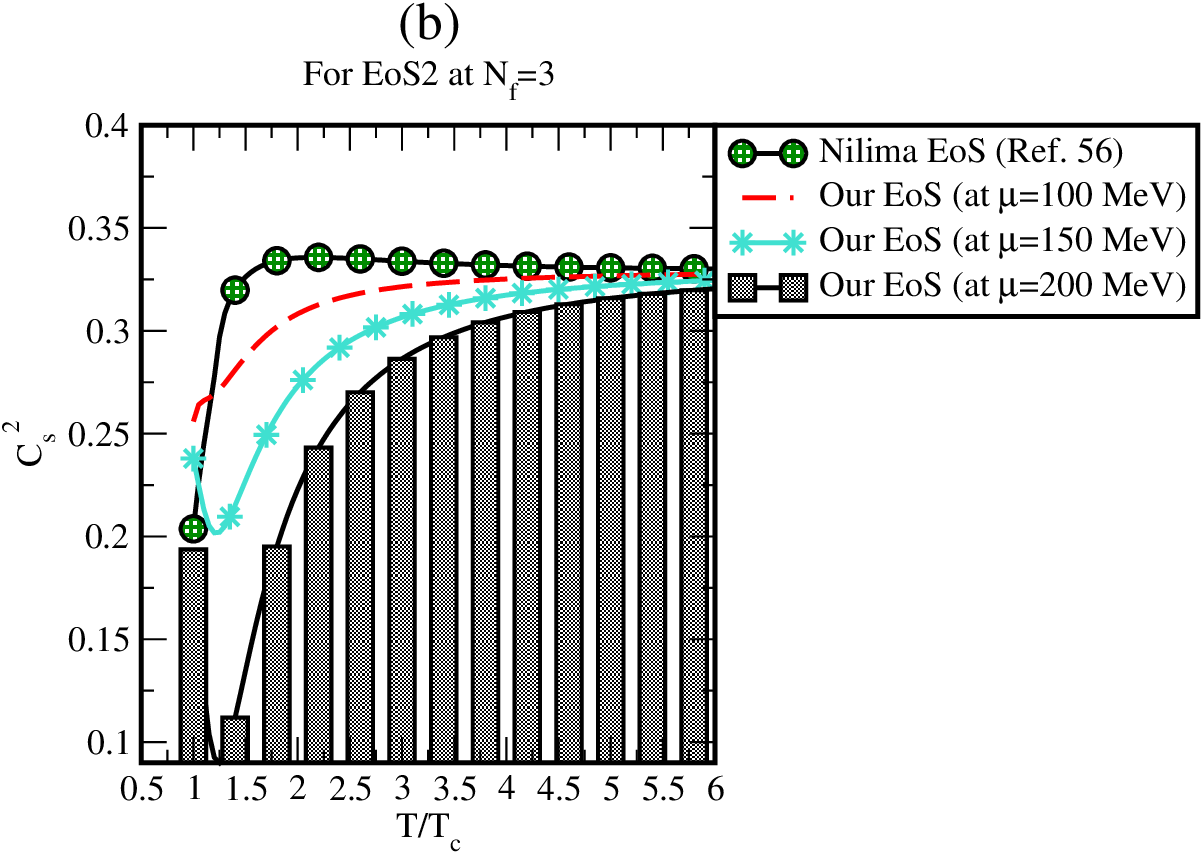}
\vspace{2cm}
\caption{Plot of $c_{s}^{2}$ as a function of $\frac{T}{T_{c}}$ for 3-flavor QGP for EoS1 (fig.5(a)) and for EoS2 (fig.5(b)).}
\label{fig.5}
\vspace{2cm}
\end{figure*}
\begin{figure*}
\includegraphics[height=5cm,width=6cm]{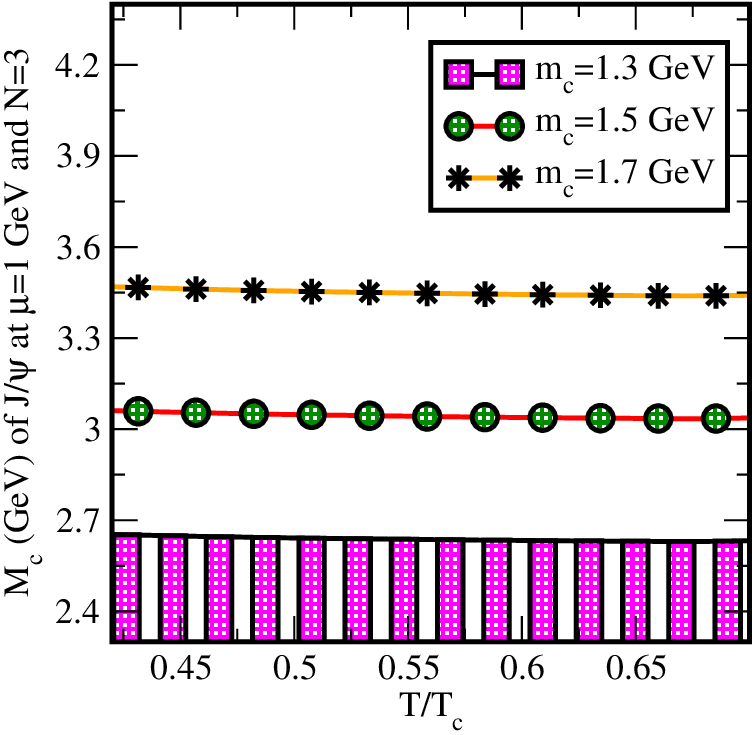}
\hspace{3mm}
\includegraphics[height=5cm,width=6cm]{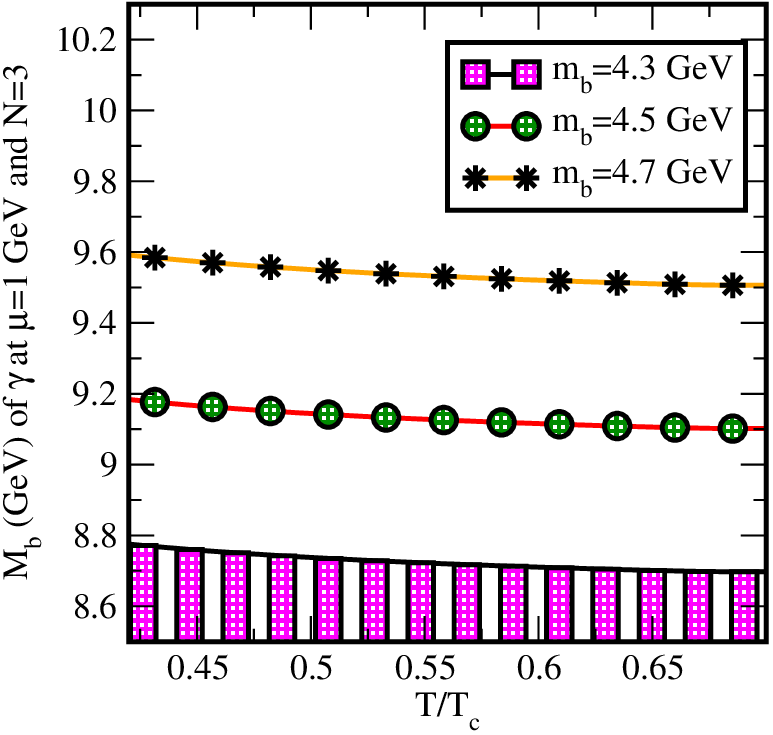}
\vspace{2cm}
\caption{Dependence of mass spectra of charmonium with $T/T_c$ (left panel) and dependence of mass spectra of bottomonium with $T/T_c$ (right panel) at $N$=$3$.}
\label{fig.6}
\vspace{2cm}
\end{figure*}

\section{The thermodynamical properties of matter with baryonic chemical potential using the formulation of equation of states (EoS)}

The equation of states (EoS) for the quark-matter is an important findings in relativistic nucleus-nucleus collisions, and the thermodynamical properties of matter are sensitive to it. The EoS which is defined as a function of plasma parameter $(\Gamma )$\cite{22} is,
\begin{eqnarray}
\label{eq43}
\epsilon_{QED}=\left [ \frac{3}{2}+\mu_{ex}(\Gamma )  \right ]nT
\end{eqnarray}
The ratio of average potential energy to average kinetic energy is known as plasma parameter. Now, let us assume that $\Gamma \ll 1$ given by,
\begin{eqnarray}
\label{eq44}
\Gamma\equiv \frac{<PE>}{<KE>}= \frac{Re[V(r,T)]}{T}
\end{eqnarray}
But after inclusion of relativistic and quantum effects, the EoS which has been obtained in the $\Gamma$ can be written as,
\begin{eqnarray}
\label{eq45}
\epsilon =\left ( 3+\mu_{ex}(\Gamma) \right )nT
\end{eqnarray}
The scaled-energy density is written as in terms of ideal contribution\cite{V.Agotiya(B.S)} is,
\begin{eqnarray}
\label{eq46}
e(\Gamma)\equiv \frac{\epsilon }{\epsilon_{SB } }=1+\frac{1}{3}\mu_{ex}(\Gamma)
\end{eqnarray}
Where, $\epsilon_{SB}$ is,
\begin{eqnarray}
\label{eq47}
\epsilon_{SB}\equiv (16+21N_{f}/2)\pi^{2}T^{4}/30
\end{eqnarray}\\
Here, $N_{f}$ is the number of flavor of quarks and gluons, and we also consider two-loop level QCD running coupling constant ($\alpha$) in $\overline{MS}$ scheme\cite{V.Agotiya(B.S)}.
\begin{figure*}
\includegraphics[height=5cm,width=6cm]{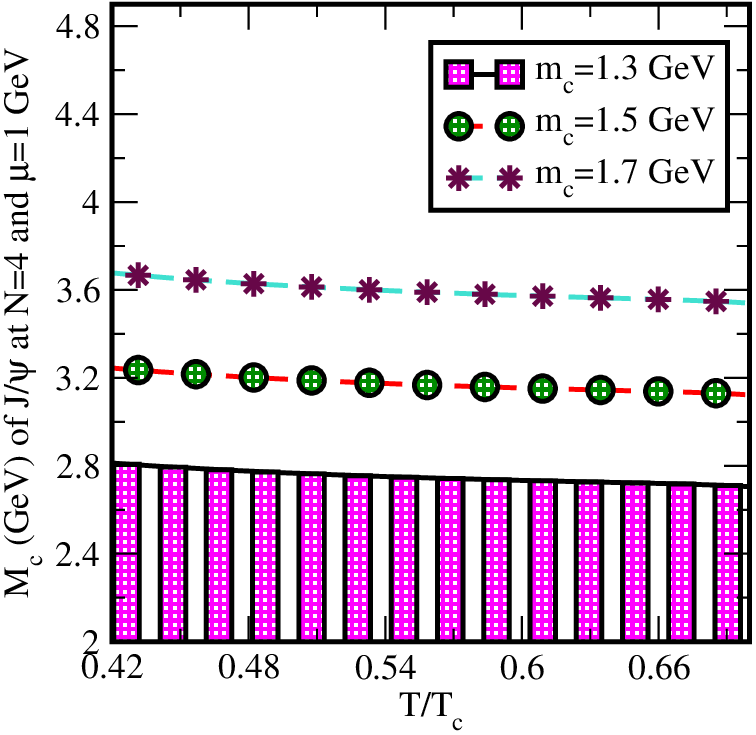}
\hspace{3mm}
\includegraphics[height=5cm,width=6cm]{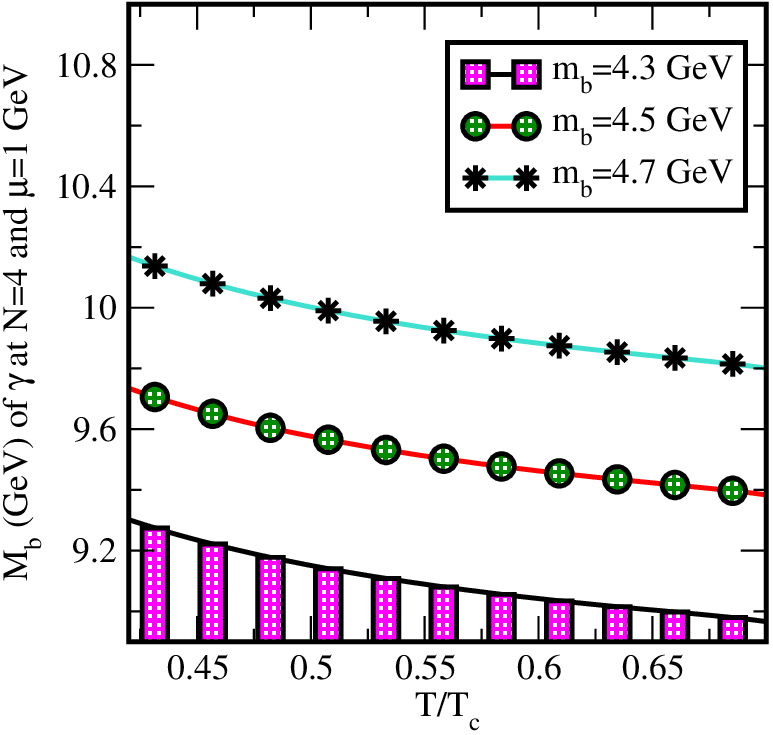}
\vspace{2cm}
\caption{Dependence of mass spectra of charmonium with $T/T_c$ (left panel) and dependence of mass spectra of bottomonium with $T/T_c$ (right panel) at $N$=$4$.}
\label{fig.7}
\vspace{3cm}
\end{figure*}
\begin{eqnarray}
\label{eq48}
g^{2}(T)\approx 2b_{0}ln\frac{\bar{\mu}}{\Lambda_{\overline{MS}} }\left ( 1+\frac{b_{1}}{2b_{0}^{2}}\frac{ln\left ( 2ln\frac{\bar{\mu}}{\Lambda_{\overline{MS}} } \right )}{ln\frac{\bar{\mu}}{\Lambda_{\overline{MS}} }} \right )^{-1}
\end{eqnarray}
Here, $b_{0}$=$\frac{33-2N_{f}}{48\pi^{2}}$ and $b_{1}$=$\frac{153-19N_{f}}{384\pi^{4}}$. In $\overline{MS}$ scheme, $\Lambda_{\overline{MS}}$ and $\bar{\mu}$ are the renormalization scale and the scale parameter respectively. And the dependency of $\Lambda_{\overline{MS}}$ is,
\begin{eqnarray}
\label{eq49}
\bar{\mu}exp(\gamma_{E}+c)=\Lambda _{\overline{MS}}(T)\nonumber\\
\Lambda _{\overline{MS}}(T)exp(\gamma_{E}+c)=4\pi\Lambda_{T}.
\end{eqnarray}
Here, $\gamma_{E}$=0.5772156 and $c$=$\frac{N_{c}-4N_{f}ln4}{22N_{c}-N_{f}}$\cite{V.Agotiya(B.S)}
After using these above relation, first we calculate the energy density $\epsilon_{T}$ from Eq.\ref{eq46} and using the thermodynamical relation,
\begin{eqnarray}
\label{eq50}
\epsilon =T\frac{dp}{dT}-P
\end{eqnarray}
We calculate the pressure as:
\begin{eqnarray}
\label{eq51}
\frac{P}{T^{4}}=\left ( \frac{P_{0}}{T_{0}}+3a_{f}\int_{T_{0}}^{T}d\tau \tau^{2}\epsilon (\Gamma (\tau ))  \right )/T^{3}
\end{eqnarray}
Here, $P_{0}$ is the pressure at some reference temperature, $T_{0}$. Now, the speed of sound is calculated by using the relation as,
\begin{eqnarray}
\label{eq52}
c_{s}^{2}=\left ( \frac{dP}{d\epsilon } \right )
\end{eqnarray}
\begin{figure*}
\centering
\includegraphics[height=7cm,width=8cm]{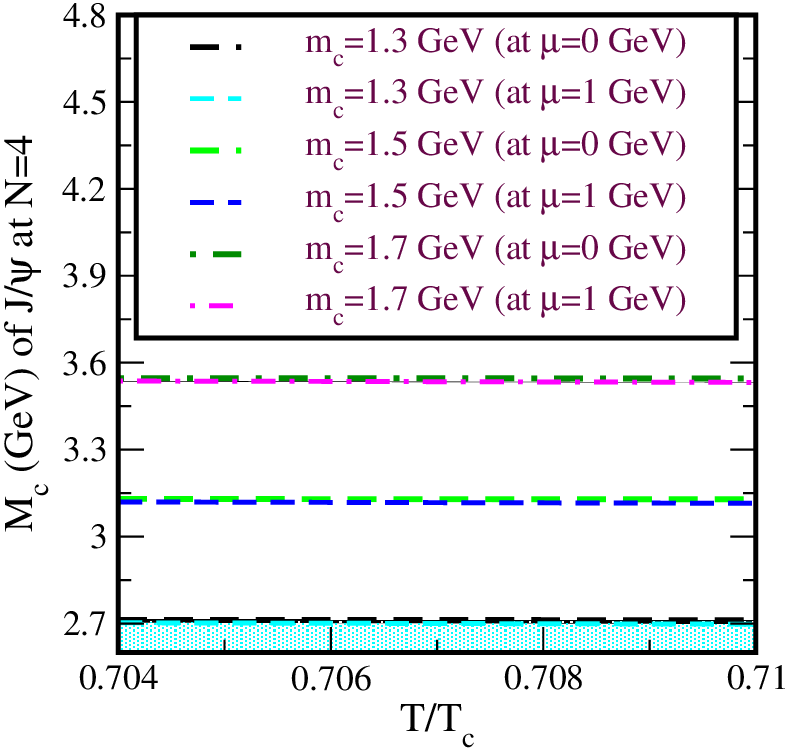}
\caption{Dependence of mass spectra of charmonium with $T/T_{c}$ at different values $\mu$ at N=4.}
\label{fig.8}
\vspace{1mm}
\end{figure*}
We know, P is the pressure and $\epsilon$ is the energy density. All these above thermodynamical properties is potential dependent, and the potential is debye mass dependent. Hence in that case, we invade the problem by trading off the dependence on baryonic chemical potential, temperature to a dependence on these thermodynamic properties of matter.

\section{Mass Spectra of Quarkonium state}

The mass spectra of heavy quarkonia can be calculated by using the relation given below:
\begin{equation}
\label{eq53}
M=2m_{Q}+ B.E.
\end{equation}
Here, mass spectra is equal to the sum of the binding energy (B.E.) and twice of the quark-antiquark mass. Substituting the values of $E_{n,l}$ in the Eq.(\ref{eq47}) we get,
\begin{multline}
\label{eq54}
M=2m_{Q}+\frac{1}{m_{Q}}\left[\frac{ \left( \frac{-m_{D}+\sqrt{m_{D}^{2}+4 m_{D}^{2}\left [ \left ( l+\frac{N-2}{2} \right )^{2}-\frac{1}{4} \right ]}}{2}-nm_{D} \right )^{2}+\left \{m_{D}^{2}\left [ \left ( l+\frac{N-2}{2} \right )^{2}-\frac{1}{4} \right ]-m_{Q}\alpha m_{D}\right \}}{2\left ( \frac{-m_{D}+\sqrt{m_{D}^{2}+4m_{D}^{2}\left [ \left ( l+\frac{N-2}{2} \right )^{2}-\frac{1}{4} \right ]}}{2}-nm_{D} \right )} \right ]^{2}\\+\frac{1}{m_{Q}}\left \{m_{D}^{2}\left [ \left ( l+\frac{N-2}{2} \right )^{2}-\frac{1}{4} \right ]-m_{Q}\alpha m_{D}\right \}
\end{multline}
Where, $m_{Q}$ is the mass of quarkonium state, N is the dimensionality number and $m_{D}$ is the QP debye mass with baryonic chemical potential.

\section{Results and Conclusion}

In this work, we have studied the properties of the quark gluon plasma by using the SUSY quantum mechanics method for solving the Schroedinger equation and hence obtained the energy eigen-value of the quarkonium states (i.e., charmonium and bottomonium) in the N-dimensional space. Thereafter, we studied the variation of binding energy of the $J/\psi$, $\Upsilon$ and this is shown in figures (\ref{fig.1}) and (\ref{fig.2}). Figure (\ref{fig.1}) shows the variation of the binding energy with temperature at $N$=$2$, $3$ and $4$ at fixed value of the baryonic chemical potential $\mu$=$300$ MeV for $J/\psi$ (left panel) and for $\Upsilon$ (right panel). It has been observed that the binding energy decreases with the temperature. We have also noticed that, as the dimensionality number increases, the variation of binding energy also increases.\\
\begin{table}[ph]
\tbl{Mass spectra for $1S$ states of Quarkonium (is in unit of GeV) at $\mu$=$1000$ MeV, $N$=$3$ and $T_{c}$=$197$ MeV (Present data related for quasi-particle Debye mass with baryonic chemical potential).}
{\begin{tabular}{@{}ccccc@{}} \toprule
$State$ & present & Shady et al & Shady & Experimental \\
& work & (2019)\cite{9} & (2016)\cite{36} & Data\cite{Tanabashi} \\ \colrule 
$J/\psi$ & 3.060 & 3.096 & 3.096 & 3.096 \\
$\Upsilon$ & 9.200 & 9.460 & 9.460 & 9.460 \\ \botrule
\end{tabular} \label{ta1}} 
\end{table}
Figure (\ref{fig.2}) shows variation of the binding energy with temperature at $\mu$=$0$, $1000$ and $2000$ MeV while the dimensionality number taken as $N$=$4$ for both $J/\psi$ (left panel) and $\Upsilon$ (right panel). With the increasing value of baryonic chemical potential $\mu$=$0$, $1000$ and $2000$ MeV, binding energy of $J/\psi$ and $\Upsilon$ is found to decrease.\\
\begin{figure*}
\centering
\includegraphics[height=6cm,width=5.5cm]{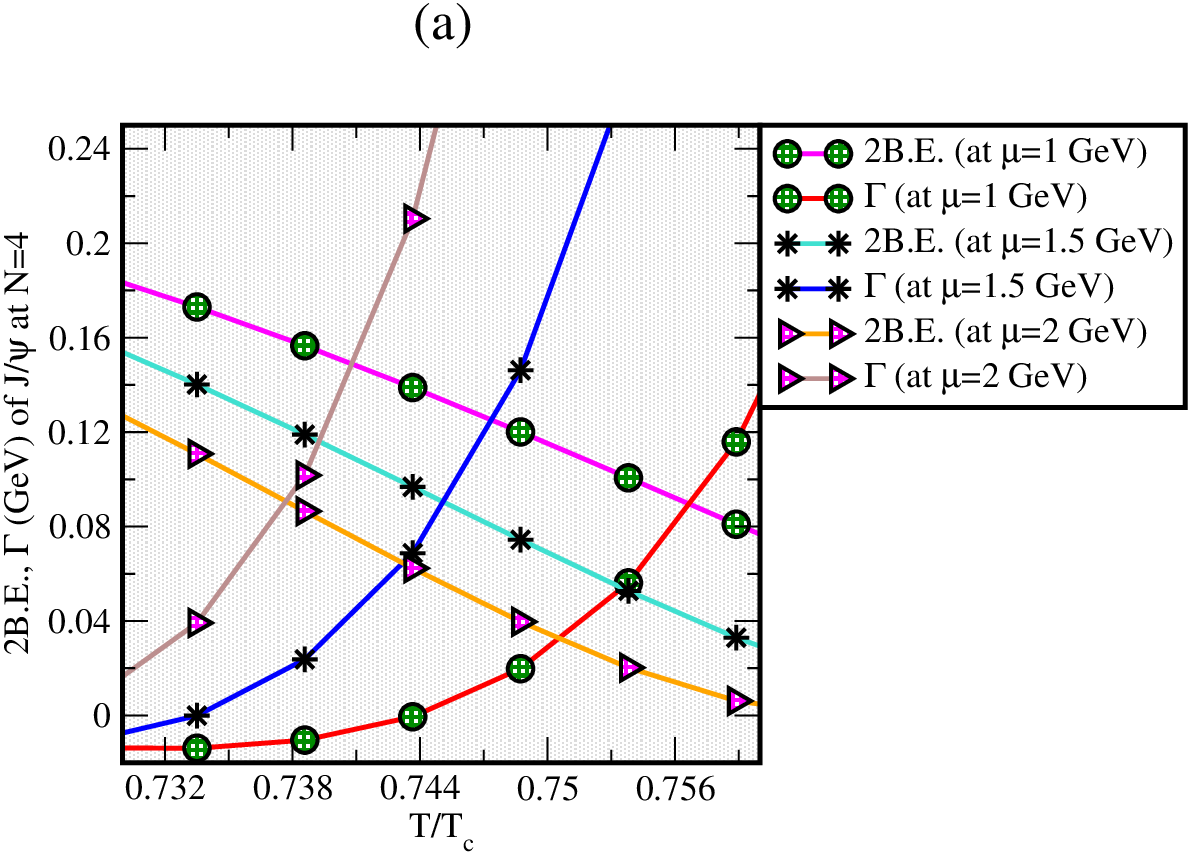}
\hspace{3mm}
\includegraphics[height=6cm,width=5.5cm]{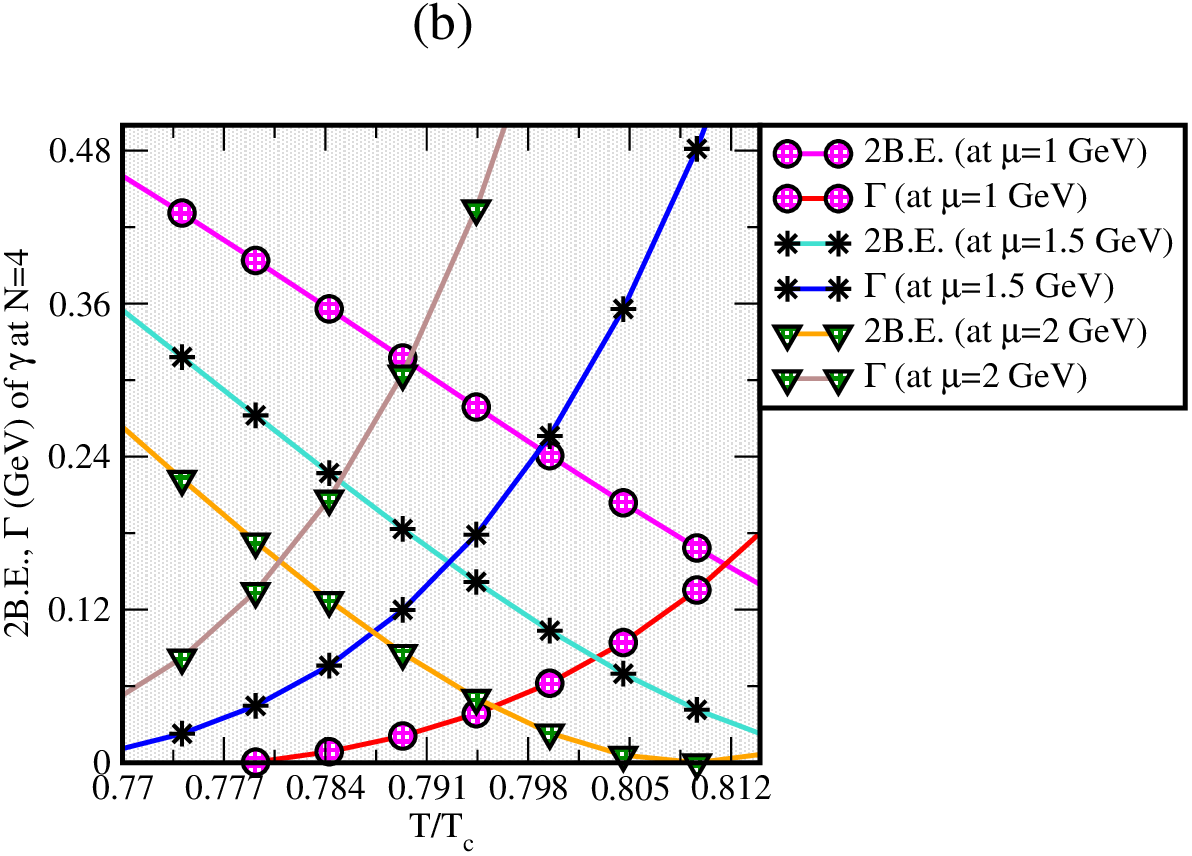}
\caption{Shows the real and imaginary binding energies of the $J/\psi$ (fig.9(a)) and $\Upsilon$ (fig.9(b)) for different values of baryonic chemical potential at $N$=4.}
\label{fig.9}
\vspace{1mm}
\end{figure*}
The thermodynamical properties of quark matter plays a significant role in the study of QGP and also provide useful information about the strange quark-matter. In figure (\ref{fig.3}) we have plotted the variation of pressure $(\frac{P}{T^{4}})$ with temperature $(T/T_{c})$ using EoS1 (fig.3(a)) and Eos2 (fig.3(b)) for 3-flavor QGP along with Nilima EoS~\cite{V.Agotiya(B.S)}. Now, energy density $\epsilon$, speed of sound $(C_{s}^{2})$, and so forth can be derived since we had obtained the pressure. In figure (\ref{fig.4}), we had plotted the energy density ($\frac{\epsilon}{T^{4}}$) with temperature $(T/T_{c})$ using Eos1 (fig.4(a)) and EoS2 (fig.4(b)) for 3-flavor QGP along with Nilima EoS~\cite{V.Agotiya(B.S)}. In figure (\ref{fig.5}), we have plotted the speed of sound $C_{s}^{2}$ with temperature ($T/T_{c}$) using Eos1 (fig.5(a)) and EoS2 (fig.5(b)) for 3-flavor QGP along with Nilima EoS~\cite{V.Agotiya(B.S)}. Our result of these thermodynamical properties of quark matter is excellently matched with the result of Nilima EoS~\cite{V.Agotiya(B.S)}. The effect of baryonic chemical potential is also observed in these thermodynamical properties of quark matter as shown in figures (\ref{fig.3}), (\ref{fig.4}) and (\ref{fig.5}). If we increase the value of baryonic chemical potential ($\mu$=$100$, $150$ and $200$) the variation of pressure, energy density and speed of sound decreases respectively.\\
The variation of mass spectra with temperature for the the $J/\psi$=$1.5$ GeV, $\Upsilon$=$4.5$ GeV has been shown by the figures (\ref{fig.6}) and figure (\ref{fig.7}). The left panel of both these figures shows the mass spectra for charmonium ($J/\psi$=$1.5$ GeV) whereas the right panel shows mass spectra for the bottomonium ($\Upsilon$=$4.5$ GeV) respectively. As we increase the value of mass of quarkonium then the variation of mass spectra also increases. The variation of mass spectra is also affected by the dimensionality number, if we increase the value of the dimensionality number the mass spectra also increases. Figure (\ref{fig.8}) shows the dependence of mass spectra of charmonium with $T/T_{c}$ at different value of $\mu$ (i.e., $\mu$= $0$ and $1$ GeV) at $N$=$4$. As we increase the value of baryonic chemical potential the variation of mass spectra decreases. In Table~\ref{ta1}, we have also calculated the values of mass spectra at $\mu$=$1000$ MeV, $N$=$3$ and $T_{c}$=$197$ MeV, and gets the value of $J/\psi$=$3.0597$ GeV (at $m_{J/\psi}$=$1.5$ GeV) and $\Upsilon$=$9.200$ GeV (at $m_{\Upsilon}$=$4.5$ GeV), and compared the values of present data with the values of Experimental Data\cite{Tanabashi}.\\
\begin{table}[ph]
\tbl{The dissociation temperature $(T_D)$ of the ${J/\psi}$ and ${\Upsilon}$ at $N$=4 has been calculated for the different values of baryonic chemical potential when the thermal width $\Gamma=2B.E$.}
{\begin{tabular}{@{}cccc@{}} \toprule
$State$ & $\mu$=1 GeV & $\mu$=1.5 GeV & $\mu$=2 GeV \\ \colrule 
$J/\psi$ & 0.7567 & 0.7450 & 0.7376 \\
$\Upsilon$ & 0.8116 & 0.7926 & 0.7808 \\ \botrule
\end{tabular} \label{ta2}} 
\end{table}
We have also calculated the dissociation temperature ($T_{D}$) for the both ground states of quarkonium. Different types of dissociation criteria has been discussed in the literature. The first criteria is that the quarkonium state should dissociate at the temperature, where $B.E.$=$T$ (for upper bound state $T_{D}$) and $B.E.$=$3(T)$ (for lower bound states $T_{D}$), hence B.E. is the binding energy of that particular quarkonium state. Here we use a more strict dissociation criteria which suggests that any quarkonium state should dissociate at temperature, where the decay width of the quarkonium state become equal to two times of binding energy. In figure (\ref{fig.9}) we have plotted the decay width and two times of binding energy of $J/\psi$ (fig.9(a)) and for $\Upsilon$ (fig.9(b)) with respect to $T/T_{c}$ at different values of baryonic chemical potential ($\mu$= $1$, $1.5$ and $2$ GeV). Now, the values of $T_{D}$ at different values of $\mu$ is given in the Table~\ref{ta2} by using the figure (\ref{fig.9}). We observed that, because of the dependency of baryonic chemical potential the values of dissociation temperature of $J/\psi$ and $\Upsilon$ is lower as compared to the values of $T_{D}$ of $J/\psi$ and $\Upsilon$ without the effect of $\mu$.\\
To conclude, the present result provide a good description for quarkonium in hot and dense media to analyze the baryon rich quark gluon plasma which is expected at Facility for Anti-proton and Ion research (FAIR) energies will be created. The form of medium modified Cornell potential procure conclusive result for binding energy, thermodynamical properties and mass spectra of the ground states of quarkonia in N-dimensional space at $N$=$4$.
Subsequently, we would enhance our study in future for the description of hybrid stars and also calculate the suppression of quarkonia in terms of baryonic chemical potential with the help of thermodynamical properties of quark matter. We will also study both the macroscopic and microscopic properties of neutron star using anomalous magnetic moment effect (AMM) in N-dimensional space. Such type of studies can be helpful in explaining the interior of the strongly magnetized neutron star.\\

\section*{Acknowledgement}
One of the author, VKA acknowledge the Science and Engineering Research Board (SERB), Project No. {\bf EEQ/2018/000181}, New Delhi for providing the financial support. We record our sincere gratitude to the people of India for their generous support for the research in basic sciences.


\begin{thebibliography}{0}
\bibitem{1} D. Teaney, J. Lauret and E. V. Shuryak, {\it Phys. Rev. Lett.} {\bf 86} (2001) 4783 [nucl-th/0011058] [INSPIRE].
\bibitem{2} D. Teaney, J. Lauret and E. Shuryak, {\it nucl-th/0110037} [INSPIRE].
\bibitem{3} P. F. Kolb and U. W. Heinz, {\it nucl-th/0305084} [INSPIRE]. 
\bibitem{4} P. Jacobs and X. N. Wang, {\it Prog. Part. Nucl. Phys.} {\bf 54} (2005) 443 [hep-ph/0405125] [INSPIRE].
\bibitem{5} P. Romatschke, {\it Int. J. Mod. Phys. E} {\bf 19} (2010) 1 [arXiv:0902.3663] [INSPIRE]
\bibitem{6} Y. Aoki, G. Endrodi, Z. Fodor, S. Katz and K. Szabo, {\it Nature} {\bf 443} (2006) 675 [hep-lat/0611014] [INSPIRE].
\bibitem{7} P . Hasenfratz and F. Karsch, {\it Phys. Lett. B} {\bf 125} 308 (1983); J. B. Kogut et al., {\it Nucl. Phys. B} {\bf 225} 93 (1983).
\bibitem{8} A. N. Ikot, B. C. Lutfuoglu, M. I. Ngwueke, M. E. Udoh, S. Zare, H. Hassan, {\it Eur. Phys. J. Plus} {\bf 131}, 419 (2016). 
\bibitem{9} M. Abu-Shady, T. A. Abdel-Karim, S. Y. Ezz-Alarab, {\it J. Egypt. Math. Soc.} {\bf 27}, 14 (2019). 
\bibitem{10} A. Suparmi, C. Cari, A. S. Husein, H. Yulian, I. K. A. Khaled, H. Luqman, E. Supriyanto, {\it in 4th International Conference on Advanced Nuclear Science Engineering} {\bf 1615}, 121–127 (2014)
\bibitem{11} H. Ciftci, R. L. Hall, N. Saad, {\it J. Phys. A} {\bf 38}, 1147 (2005)
\bibitem{12} A. Sommerfeld, {\it Wave-Mechanics}, London, UK, (1930).
\bibitem{13} D. J. Grifths, Introduction to Quantum Mechanics, Pearson Prentice Hall, {\it Upper Saddle River}, NJ, USA, {\bf 2nd edition}, (2005). 
\bibitem{14} R. P. Feynman and A. R. Hibbs, Quantum Mechanics and Path Integrals, McGrawHill, New York, NY, USA, (1965).
\bibitem{15} T. Das, {\it Electronic Journal of Teoretical Physics}, {\bf 13}, 207, (2016). 
\bibitem{16} G. Chen, {\it Zeitschrif fur Naturforschung}, {\bf 59a}, 875, (2004).
\bibitem{17} E. M. Khokha, M. Abu-Shady, and T. A. Abdel-Karim, {\it International Journal of Teoretical and Applied Mathematics}, {\bf 2}, 86 (2016).
\bibitem{18} S. Ikhdair and R. Sever, {\it Journal of Molecular Structure}, {\bf 855}, 13 (2008).
\bibitem{19} R. Kumar and F. Chand, {\it Communications in Teoretical Physics}, {\bf 59}, 528–532 (2013).
\bibitem{20} S. M. Kuchin and N. V. Maksimenko, {\it Universal Journal of Physics and Application}, {\bf 7}, 295 (2013).
\bibitem{21} U. Kakade, B. K. Patra {\it Phys. Rev. C} {\bf 92}, 024901 (2015).
\bibitem{22} V. K. Agotiya, V. Chandra and B. K. Patra, {\it Phys. Rev. C} {\bf 80}, 025210 (2009).
\bibitem{23} M. Abu. Shady, {\it Europian Phys. Journal Plus} {\bf 134}, 321 (2019).
\bibitem{24} D. Kharzeev, L. McLerranand, and H. Satz, {\it Physics Letters B}, {\bf 356}, 349 (1995).
\bibitem{25} H. Satz, Charm and Beauty in a Hot Environment, {\it BI-TP 2006/06}, https://arxiv.org/abs/hep-ph/0602245 (2006).
\bibitem{26} A. N. Ikot, S. E. Etuk, B. H. Yazarloo, S. Zarrinkamar, H. Hassanabadi, {\it Few-Body Syst.}, {\bf 56}, 41 (2015).
\bibitem{27} A. N. Ikot, H. Hassanabdi, H. P. Obong, Y. E. Chad-Umoren, C. N. Isonguyo, B.H. Yazarloo, {\it Chin. Phys. B} {\bf 23}, 120303 (2014). 
\bibitem{28} A. N. Ikot, H. Hassanabdi, E. Maghsoodi, S. Zarrinkamar, N. Salehi, {\it Phys. Part. Nucl. Lett.}, {\bf 11}, 443 (2014). 
\bibitem{29} A. N. Ikot, H. P. Obong, H. Hassanabadi, N. Salehi, O. S. Thomas, {\it Ind. J. Phys.} {\bf 89}, 649 (2015). 
\bibitem{30} C. S. Jia, J. W. Dai, L. H. Zhang, J. Y. Liu, X. L. Peng, {\it Phys. Lett. A} {\bf 379}, 132 (2015).
\bibitem{31} A. Suprami, C. Cari, B. N. Pratiwi, {\it J. Phys. Conf. Ser.} {\bf 710}, 012026 (2016).
\bibitem{32} C. S. Jia, T. Chen, S. He, {\it Phys. Lett. A} {\bf 377}, 682 (2013).
\bibitem{33} M. Abu. Shady, Sh. Y. Ezz. Alarab, {\it Few Body Systems} {\bf 60}, 66 (2019).
\bibitem{34} E. witten, {\it Nucl phys B}, {\bf 188}, 513 (1981).
\bibitem{35} T. Das, A. Arda, {\it Adv. High Energy Phys.} {\bf 2015}, 137038 (2015).
\bibitem{36} M. Abu-shady, {\it Inter. J. Appl. Math. Theor. Phys.} {\bf 2}, 16 (2016).
\bibitem{37} E. Witten, {\it Nucl. Phys. B} {\bf 188}, 513 (1981).
\bibitem{38} F. Cooper, A. Khare, U. Sukhatme, {\it Phys. Rep.} {\bf 251}, 267 (1995).
\bibitem{39} H. Motavali, A. Rostami, {\it Prog. Electromagn. Res. C} {\bf 1}, 131 (2008).
\bibitem{40} A. Mocsy and P. Petreczky, {\it Physics Review Letters} {\bf 99}, n0-21, id-211602 (2007).
\bibitem{41} M. Laine, O. Philipsen and M. Tassler, {\it Journal of high energy physics} {\bf 09}, pp.066 (2007).
\bibitem{ARebhan} A. K. Rebhan et al, {\it Phys. Rev. D} {\bf 48}  R3967 (1993).
\bibitem{EBraaten} E. Braaten and A. Nieto, {\it Phys. Rev. Lett.} {\bf 73}  2402 (1994). 
\bibitem{PArnold} P. Arnold and C. Zhai, {\it Phys. Rev. D} {\bf 51} 1906 (1995).
\bibitem{KKajantie} K. Kajantie, M. Laine, J. Peisa, A. Rajantie, K. Rummukainen, and M. E. Shaposhnikov, {\it Phys.Rev.Lett.} {\bf 79} 3130  (1997). 
\bibitem{SNadkarni} S. Nadkarni, {\it Phys. Rev. D} {\bf 33}, 3738 (1986); {\bf 34} 3904 (1986).
\bibitem{YBurnier} Y. Burnier and A. Rothkopf, {\it Phys. Lett. B} {\bf 753} 232-236 (2016).
\bibitem{VGoloviznin} V. Goloviznin and H. Satz, {\it Z.Phys. C} {\bf 57} 671 (1993).
\bibitem{peshier} A. Peshier, B. Kampfer, O. P. Pavlenko, and G. Soff, {\it Phys. Rev. D} {\bf 54} 2399 (1996).
\bibitem{MDElia} M. D'Elia, A. Di Giacomo, and E. Meggiolaro, {\it Phys. Lett. B} {\bf 408} 315 (1997); {\it Phys. Rev.D} {\bf 67} 114504 (2003).
\bibitem{ADumitru} A. Dumitru and R. D. Pisarski, {\it Phys.Lett.B} {\bf 525} 95 (2002).
\bibitem{VChandra7} V. Chandra, R. Kumar, V. Ravishankar, {\it Phys. Rev. C} {\bf 76} 069904 (2007).
\bibitem{VChandra9} V. Chandra, A. Ranjan, V. Ravishankar, {\it Eur. Phys. J. A} {\bf 40} 109-117 (2009).
\bibitem{PKSrivastava} P. K. Srivastava, S. K. Tiwari, and C. P. Singh, {\it Phys. Rev. D} {\bf 82} 014023 (2010).
\bibitem{Ikot18} A. N. Ikot, E. O. Chukwocha, M. C. Onyeaju, C. A. Onate, B. C. Ita, and M. E. Udoh, {\it Pramana Journal of Physics} {\bf 90}, 22-34 (2018).
\bibitem{V.Agotiya(B.S)} I. Nilima and V. K. Agotiya, {\it Advances in high energy physics}, Article id-8965413, (2018).
\bibitem{DEbert}  D. Ebert, R. N. Faustov, V. O. Galkin, {\it Phy. Rev. D} {\bf 67}, 014027 (2003).
\bibitem{TDas} T. Das, {\it EJTP} {\bf 13}, 207 (2016).
\bibitem{SRoy} D. K Choudhury, {\it Can. J. Phys.} {\bf 94}, 1282 (2016).
\bibitem{Tanabashi} M. Tanabashi, C. Carone, T. G. Trippe and C. G. Wohl, Particle Data Group {\it Phys. Rev. D} {\bf 98}, 546-548 (2018).
\bibitem{Okorie} U. S. Okorie, E. E. Ibekwe, A. N. Ikot, M. C. Onyeaju, and E. O. Chukwuocha, {\it Journal of the korea, Physical Society}, {\bf 73}, 1210-1218 (2018).
\bibitem{Agotiyaprd} V. K. Agotiya, V. Chandra, M. Y. Jamal and I. Nilima, {\it Phys. Rev. D} {\bf 94}, 094006 (2016).
\end{thebibliography}
\end{document}